\documentclass[a4paper,10pt]{article}
\usepackage[utf8x]{inputenc}
\begin{document}
 \title{Neutrino production states and NSI}% Force line breaks with \\
\author{R. Szafron\footnote{robert.szafron@us.edu.pl},  M. Zralek\footnote{marek.zralek@us.edu.pl}\\
 Institute of Physics, University of Silesia\\
ul. Uniwersytecka 4, PL-40007 Katowice, Poland}

\date{\today}
\maketitle
\begin{abstract}
 The problems of  neutrino production states, prepared to the oscillation process, in the case of non standard interactions, are briefly discussed. Quantum neutrino states are determined from the dynamics of a production process.  We show, that even in models where only left-handed neutrinos are introduced, the standard adopted procedure is valid only approximately. Entanglement between neutrino masses or between masses, flavour and spins cause, that their quantum states are mixed. 
 \end{abstract}
 
\section{Introduction}
 Without  the Non-Standard Interaction (NSI) the existing theory of neutrino oscillation works very well. In the Standard Model (SM) neutrinos interact  through the  left-handed vector currents. Then the relativistic neutrinos (antineutrinos) have only negative (positive) helicity.  The  Maki-Nakagawa-Sakata-Pontecorvo (MNSP) mixing matrix present in the model and definite helicity cause  that neutrinos are produced (and detected) in a pure Quantum Mechanical (QM) states. The neutrino helicity is only a "spectator", it does not change at any phase of an oscillation process, even if a flavour is changing.
\begin{eqnarray}
\label{neutrin SM states} 
\vert\nu_{\alpha},\downarrow\rangle =\sum_{i}U_{\alpha,i}^{*}\vert\nu_{i},\downarrow\rangle,  \ \ \  \vert\overline{\nu}_{\alpha},\uparrow\rangle =\sum_{i}U_{\alpha,i}\vert\overline{\nu}_{i},\uparrow\rangle.
\end{eqnarray}

The fact that the NSI can change the process of neutrino oscillation for the first time was observed in  \cite{GrossmanNSI},  and then has been considered in various physical processes (see e.g. \cite{NSIo1, NSIo2,NSIo3} and references therein).  In all this approaches instead of  the states (\ref{neutrin SM states}) the new one,  connected in some way with dynamics of the processes are defined separately for production (s) and detection (d) 
\begin{eqnarray}
\label{new production and detection states}
\vert\nu_{\alpha}^{s}\rangle =\sum_{i}U_{\alpha,i}^{s}\vert\nu_{i}\rangle,  \ \ \  \vert\nu_{\beta}^{d}\rangle =\sum_{i}U_{\beta,i}^{d}\vert\nu_{i}\rangle,
\end{eqnarray}
where the mixing matrices $U_{\alpha,i}^{p}$ and $U_{\beta,i}^{d}$ depends on the NSI 
\begin{eqnarray}
\label{new production and detection mixing matrices}
\vert U_{\alpha,i}^{s}\vert^{2} \propto \vert \langle \nu_{i}; f_{P}\vert H^{P}\vert l_{\alpha}; i_{P}\rangle\vert^{2},  \  \  \ \vert U_{\beta,i}^{d}\vert^{2} \propto \vert \langle l_{\beta}; f_{D}\vert H^{D}\vert \nu_{i}; i_{D}\rangle\vert^{2}.
\end{eqnarray}
with $H^{P}$ $(H^{D})$ being the production (detection) Hamiltonian.

Such description of the oscillation phenomena in frame of NSI is widely applicable (see e.g. \cite{NSI1, NSI2, NSIS}). However, it was observed before (see Ref.\cite{OSZ, prog}), that such approach is valid only approximately. There are models which require an extension of the traditional approach presented above.  
There can be found many examples of such theories in literature, where such an extension is necessary. Let us mention only a few of them: models with two Higgs doublets\cite{2HDM1}, Zee-Babu model\cite{Zee-Babu}, models with extended symmetry group\cite{331,LR} or supersymmetry\cite{SUSYRv}.
  
   In this work we would like to reconsider once more the impact of NSI for production process. We concentrate on processes which are interesting from the practical point of view as for example; pion decay used in the present accelerator experiments, nuclei beta decay for planed future beta beams, and muon decay for future neutrino factories. 
   
   We also discuss the case, where only left-handed neutrino fields enter a model and relativistic neutrinos (antineutrinos) always have negative (positive) helicity.  Then the difference between conventional approach (pure neutrino state) and the real one (mixed neutrino state) are not connected with obvious mixing of states caused by the different neutrinos helicities .
   
    As we will see in the next Chapter, the mixed state, if appear, is connected with subtle things of an entanglement between neutrinos or antineutrino  in different mass states (in the muon decay case, where neutrino-antineutrino appears) or between masses, flavour and spins, when only one neutrino is produced (pion decay, beta decay).

\section{Quantum state of neutrinos produced  in the processes with NSI}
The decay of pions, nuclei, or muons are in the lowest order described by the general  d=6 effective Lagrangian 
\begin{eqnarray}\label{eq:NSIG}
{\cal L}_{\rm NSI} = -2\sqrt{2}{G_F}
\sum_{\Delta=S,V,T \atop \varepsilon,\eta=L,R}
\varepsilon^{\Delta}_{\varepsilon,\eta} \left(
\overline{\psi}_{a}\Gamma^\Delta P_{\varepsilon} \psi_{b} \right) \left(
\overline{\psi}_{c}\Gamma_{\Delta} P_{\eta} \psi_{d} \right)\ +h.c. ,
\end{eqnarray}
 where the field $\psi_{a} (\psi_{b})$ describes the produced neutrino (antineutrino).  In many extensions of the Standard Model (MS) only left-handed neutrino fields $\psi_{a,L}$ are present, in such a case the  Lagrangian (Eq.(\ref {eq:NSIG})) simplifies (five parameters; $\varepsilon^{S}_{R,R(L)}, \varepsilon^{V}_{R,R(L)} $ and $ \varepsilon^{T}_{R,L}$ are then not vanishing). If two left-handed neutrinos  appear (e.q. in addition $d=\nu$, like in the muon decay) then only one scalar and one vector term remains ($\varepsilon^{S}_{R,L},\varepsilon^{V}_{L,L}\neq 0$). Then the Lagrangian (\ref {eq:NSIG}), by the Fierz rearrangement theorem (see e.q.\cite{Fierz}), has the form similar to the one used for studding the NSI in matter (see e.g. \cite{Wintlect, ogr1, ogr2}), 
\begin{eqnarray}\label{eq:NSI}
{\cal L}_{\rm NSI} = -2\sqrt{2}{G_F}
\sum_{\alpha,\beta =e,\mu,\tau \atop C=L,R}
\varepsilon^{ff'C}_{\alpha\beta} \left(
\overline{\nu}_\alpha\gamma^\mu P_{L} \nu_\beta \right) \left(
\overline{f}\gamma_\mu P_{C} f' \right)\ + h.c. ,
\end{eqnarray}
where $\varepsilon^{V}_{L,L} \Rightarrow -\varepsilon^{f,f^{'} L}_{\alpha,\beta}$,  $\varepsilon^{S}_{R,L} \Rightarrow 2 \varepsilon^{f,f^{'} R}_{\alpha,\beta}$ and $ \psi_{a} \Rightarrow \nu_{\alpha}, \psi_{b} \Rightarrow f^{'} , \psi_{c} \Rightarrow  f, \psi_{d} \Rightarrow \nu_{\beta}$. 
On what follows we will consider this form of the NSI.

Despite that the main effects of this interaction is to modify the neutrino oscillations in matter, it is also claimed that NSI modifies the production and detection states such that those states are given by (not normalised) see e.g. \cite{GrossmanNSI, Wintlect, pureCP}:
\begin{eqnarray}\label{eq:SD}
|\nu^s_\alpha \rangle & = &|\nu_\alpha \rangle +
\sum_{\beta=e,\mu,\tau} \varepsilon^s_{\alpha\beta}
|\nu_\beta\rangle   \ , \\\label{SD2} \langle \nu^d_\beta| & = &  \langle
\nu_\beta | + \sum_{\alpha=e,\mu,\tau} \varepsilon^d_{\alpha \beta}
\langle  \nu_\alpha  | \ ,
\end{eqnarray}
where the parameters $ \varepsilon^{s(d)}_{\alpha \beta}$ are connected with the complex coefficients $\varepsilon^{ff'C}_{\alpha\beta}$ in the way which depend on the given production (s) and detection (d) process.

We would like to reconsider once more  the impact of NSI given by (\ref{eq:NSI}) for the production process. 
In order to calculate the state of neutrino produced in some reaction we  write the density matrix describing a final state of this reaction  \cite{OSZ,prog} and then, using this final quantum state, we take the partial trace over unobserved particles in order to obtain the state of neutrino which we are interested in.  
We will also work in a mass base rather than in flavour base. The mass base is convenient because it is well defined and suitable for considerations of the oscillation phenomenon. To move from flavour to mass base we assume as usual that flavour neutrino is given as a  linear combination of mass eigenstates i.e.

\begin{eqnarray}\label{eq:MM}
\nu_\alpha =\sum_i U_{\alpha i} \nu_i
\end{eqnarray}
where $U$ is some unitary matrix, in SM it is just the MNSP mixing matrix. Generally left- and right-handed neutrino fields can have, depending on the neutrino mass model, different mixing matrices, but in our case  Eq. (\ref{eq:MM}) is sufficient. 

The modification  of initial and final state given by eq. (\ref{eq:SD}) and (\ref{SD2}) assumes that the all spin amplitudes are added coherently. This is not true in general for NSI (\ref{eq:NSI}). Therefore we will consider both cases i.e. coherent and incoherent NSI contribution to the production process. Furthermore the situation gets more complicated with two neutrinos in final state, therefore we will consider a muon decay as an example of production process 
\begin{eqnarray}\label{eq:MD}
 \mu^{-} \rightarrow e^{-}  \overline{\nu}_i  \nu_j,
\end{eqnarray}
but to simplify the calculation we assume that neutrinos are Dirac particles. Now in (\ref{eq:MD}) the "i" and "j" indices indicate the neutrino mass. 

If neutrinos are left-handed the most general effective Lagrangian (with the SM part included) in the mass base for the  process (\ref{eq:MD}), has the form
\begin{eqnarray}\label{eq:muon}
{\cal L}_{\rm \mu} = -2\sqrt{2}{G_F}\left[ g^S_{i j} \left(
\overline{\nu}_i P_{R}  e \right) \left(
\overline{\mu} P_{L} \nu_j\right)  + g^V_{i j}  \left(
\overline{\nu}_i\gamma^\alpha P_{L}  e \right)  \left(
\overline{\mu}  \gamma_\alpha P_{L}\nu_j \right) \right]+ h.c.
\end{eqnarray}
Taking into account the  SM part we should replace the  $\varepsilon$'s factors defined by eq. (\ref{eq:NSI}) by
\begin{eqnarray}\label{eq:IS}
 \varepsilon^{ \mu e L}_{\alpha \beta} \Rightarrow G_{\alpha \beta} = \delta_{\alpha e} \delta_{\mu \beta} + \varepsilon^{\mu e L}_{\alpha \beta},
\end{eqnarray} 
then the relation between Lagrangians (\ref{eq:NSI}) and (\ref{eq:muon}) are given by
\begin{eqnarray}\label{relations 1}
g^V_{ij}= -\sum_{\alpha, \beta}U^*_{\alpha i}\varepsilon^{\mu e L}_{\alpha\beta} U_{\beta j} - U^{*}_{e i} U_{\mu j},
\end{eqnarray}
and
\begin{eqnarray}\label{relations 2}
g^S_{ij}=  2 \sum_{\alpha, \beta}U^*_{\alpha i}\varepsilon^{\mu e R}_{\alpha\beta} U_{\beta j}.
\end{eqnarray}

We will consider two cases. First we assume that  in (\ref{eq:NSI}) only one term C = L appears, so $\varepsilon^{\mu e L}_{\alpha\beta}\neq 0$ and $\varepsilon^{\mu e R}_{\alpha\beta}=0$. In this case the NSI contribute to the same helicity amplitude as $W^\pm$ boson exchange in SM (coherent case). The density matrix which describe the antineutrino - neutrino state produced in muon decay (\ref{eq:MD}) in the mass-helicity base (it does not depend  on an initial muons  polarisation, and whether  the final electrons polarisation is measured or not)   has then the form
 \begin{eqnarray}\label{eq:DM for 2neutrinos}
\varrho(i,j;k,l) = \chi^{*}_{i,j} \chi_{k,l},
\end{eqnarray}
where $ \chi_{i,j}= \frac{g^{V}_{i j}}{N}$ and $N = \sqrt{ \sum_{ i,j} \mid g^{V}_{ij}\mid^{2} }$. For the left-handed and to a good approximation massless neutrinos, their helicities are unambiguously determined, and therefore in formula (\ref{eq:DM for 2neutrinos})  their are not indicated explicitly. Such density matrix  describes  pure QM two neutrinos state ($Tr(\varrho^{2})=1$), which can be written in the form 
\begin{eqnarray}\label{eq:2 neutrino state}
 |\overline{\nu}' \nu \rangle  =\sum_{i,j} (\chi_{ij})^{*}|\overline{\nu}_i\nu_j  \rangle =\frac{1}{N} (| \overline{\nu}_e \nu_\mu \rangle +\sum_{\alpha,\beta} (\varepsilon^{ \mu e L}_{\alpha \beta})^{*} |\overline{\nu}_\alpha \nu_\beta \rangle) .
\end{eqnarray} 
In order to get one particle state, we must take the partial trace over the  second particle degrees of freedom ($\varrho^{\overline{\nu}}(i;k) = \sum_{j} \varrho(i,j;k,j)$) for antineutrinos and over the first mass indices ($\varrho^{\nu}(j;l) = \sum_{i} \varrho(i,j;i,l)$) for neutrinos and we obtain

\begin{eqnarray}\label{eq:1czastka}
 \varrho^{\overline{\nu}}(i;k) = \frac{(U^{†} GG^{†} U)^{*}_{i k}}{Tr (G G^{†})}, \ \  \varrho^{\nu}(j;l) = \frac{(U^{†} G^{†}G U)_{j l}}{Tr (G^{†}G )},
\end{eqnarray} 
where U and G matrices are defined in Eqs (\ref{eq:MM}) and (\ref{eq:IS}) respectively.
Generally such states are not pure, because
\begin{eqnarray} \label{eq:Trro2}
Tr(\varrho^{2}) = \frac{Tr[G^{†} G G^{†} G]}{(Tr[G^{†} G])^{2}} \neq 1.
\end{eqnarray}
 Only for very specific NSI couplings such that $Tr(A^{2}) = (TrA)^2$ with $A = G^{\dagger} G$ both neutrino states produced in muon decay are pure. In general these state can not be written as a pure state because the NSI caused the entanglement of neutrino and antineutrino produced in muon decay. Only if the initial state (\ref{eq:2 neutrino state}) is factorisable then the produced neutrino can be described as a pure quantum mechanical state. If we look at the models beyond the SM, then  the parameters $\varepsilon_{\alpha \beta}^{f f' C}$ are usually parametrised in the way
\begin{eqnarray}
\varepsilon_{\alpha \beta}^{f f' C} = \frac{h_{\alpha f} h_{\beta f'}^*}{\Lambda^2},
\end{eqnarray}
where $h_{\alpha f}$ and $\Lambda$ are some couplings and a scale of the New Physics. Unfortunately this not guarantee that the state (\ref{eq:1czastka}) will be pure. This will be the case if $h_{\alpha e}=\delta_{\alpha e}$ or $h_{\alpha \mu}=\delta_{\alpha \mu}$.  In this situation the state (\ref{eq:SD}) will be the correct one. In general the difference between (\ref{eq:SD}) and (\ref{eq:IS}) is only of order $\varepsilon^2$ so then with the present bounds on NSI \cite{ogr1,ogr2} the error made by taking a pure state instead of the density matrix is very small and the standard approach is justified, but it must be remembered that (\ref{eq:SD}) is only the first order approximation. This situation is typical in reactions which involve more than one neutrino in the final state and the NSI giving a coherent contribution.

Now we consider a general case. Let us assume that NSI contributes to the scalar and vector part in (\ref{eq:muon}) so $\varepsilon^{e\mu R}_{\alpha\beta} \neq 0$ and $\varepsilon^{e\mu L}_{\alpha\beta} \neq 0$. 
Such interactions can, for example, appear in theories with charged scalar boson like models with more than one Higgs doublets \cite{2hdm}. 
 The vector and scalar terms from (\ref{eq:muon}) contribute to the same helicity muon decay amplitude (if $m_{e} \neq 0$ once more the NSI contributes coherently) but with different kinematical factors giving finally the density matrix for muon neutrino
\begin{eqnarray}\label{eq:GDM}
 \varrho  &=& \frac{1}{\overline{N}} \left( B f(E)+  C g(E) - 2 Re[(g^{V})^{T} (g^{S})^{*}] (\frac{m_{e}}{E}) h(E) \right) ,
\end{eqnarray}
where $f(E)=6M-8E+0(\frac{m_{e}}{E})$, $g(E)=3(M-2E) +0(\frac{m_{e}}{E})$ and $h(E) $ are functions depending on neutrino energy $E$ in muon rest frame and muon mass $M$. The matrices $A$ and $B$ are defined in the way $B = (g^{V})^{T} (g^{V})^{*}$  and $C =(g^{S})^{T} (g^{S})^{*}$. Such density matrix is linear in NSI parameters (third term in (\ref{eq:GDM})), but unfortunately this term is proportional to small factor  ($\frac{m_{e}}{E}$) and therefore we neglect it. Then the normalising factor is given by 
\begin{eqnarray}\label{eq:Nor}
\overline{N} = f(E) Tr B + g(E) Tr C .
\end{eqnarray}
When neglecting the electron mass $\varepsilon_{\alpha \beta}^{f f' L}$ and $\varepsilon_{\alpha \beta}^{f f' R}$ contribute to the different helicities amplitudes (incoherent case). State given by (\ref{eq:GDM}) is mixed, $Tr(\varrho^{2})$ is given by
\begin{eqnarray}\label{eq:Trrosq}
Tr(\varrho^{2}) = \frac{1}{\overline{N}^{2}} (f^{2}(E) Tr (B^{2}) + g^{2}(E) Tr(C^{2}) + 2 f(E) g(E) Tr(B C),
\end{eqnarray}
To get neutrino in a pure state for any value of neutrino energy, couplings should satisfy very restrictive conditions $Tr(B^{2}) = (TrB)^2$, $Tr(C^{2}) = (TrC)^2$ and $Tr(B C) = Tr(B) Tr(C)$,  which do not follows from any symmetry of the  considered models.
 
We see that by taking a pure state instead of the density matrix, as an initial neutrino state we overestimate the impact of NSI because it is of order $\varepsilon^2$ at most and there is no way to obtain the state (\ref{eq:SD}) as some kind of approximation of formula (\ref{eq:GDM}). In (\ref{eq:GDM}) there is also term linear in $\varepsilon$ but it is proportional to the small ratio $\frac{m_{e}}{E}$ and we neglect it. 

What can we say about the quantum states, when only one neutrino is produced, as in the pions decay  or beta decay of nuclei? As we discussed previously  - as many as five operators describe production of the left-handed neutrinos, but there is no mass entanglement between them. As as example let us take effective Lagrangian for antineutron beta decay without tensor operator

\begin{eqnarray}\label{eq:beta decay}
{\cal L}_{\rm \mu} = -2\sqrt{2}{G_F}\left[ g^{S_{L,R}}_{i} \left(
\overline{\nu}_i P_{R}  e \right) \left(
\overline{n} P_{L,R} \ p\right)  + g^{V_{L,R}}_{i}  \left(
\overline{\nu}_i\gamma^\alpha P_{L}  e \right)  \left(
\overline{n}  \gamma_\alpha P_{L,R} \ p\right) \right] + h.c.
\end{eqnarray}
If the coupling have similar structure as  before (e.g.  $g^{S_{L}}_{i} = \sum_{\alpha} g^{S_{L}}_{\alpha} U^{*}_{\alpha i}$) then the state of produced neutrino is once more mixed which is caused by the decay amplitudes
\begin{eqnarray}\label{eq:amplitudes for beta decay}
f(\lambda_{n},\lambda_{p},\lambda_{e}, \nu_i) =\sum_{\alpha,\Delta} U^{*}_{\alpha i}g^{\Delta}_{\alpha} f_{\Delta}(\lambda_{n},\lambda_{p},\lambda_{e}).
\end{eqnarray}
Where $f_{\Delta}(\lambda_{n},\lambda_{p},\lambda_{e})$ are  the amplitudes calculated for a given operator   $\Delta= S_{L,R}, V_{L,R}$. The state will be pure only if all $f_{\Delta}$'s for any spin configuration of accompanying particle are equal (it can not happened in practice) or the couplings $g^{\Delta}_{\alpha}$ do not depend on the operators $\Delta$ (strong constrain). Once again, we have kind of entanglement, in this case, between neutrino masses, flavor structure and spin projections.

We considered the NSI  with left - handed Dirac fields only, as the possible values of the right-handed couplings are strongly constrained, especially with the requirement of gauge symmetry (see e.g \cite{br}). In models which accept right-handed chiral fields, neutrinos  (antineutrinos) with a positive (negative) helicities are produced and their states are obviously mixed. 

 \section{Conclusions}
 Quantum state of neutrinos, which are produced in the present and future oscillation experiments was re-discussed. It was shown that these states depend in a crucial way on the details of the production mechanism beyond the Standard Model. Assumption that they are always pure quantum states is incorrect. Even when the helicity of neutrinos is clearly defined, their mass states and flavours remain entangled, causing that the full state of a single neutrino is mixed. Sometimes the effect can depend linearly on the parameters of New Physics and when one is searching for such an effects in neutrino oscillations must keep this in mind.  
 
\section*{Acknowledgements}
This work has been supported by the Ministry of Science under grant
No. N N202 064936 and by the European Community Marie-Curie Research
Training Network under contract MRTN-CT-2006-035505.
R.S. acknowledges a scholarship from the UPGOW project cofinanced by the European Social Fund.

\end{document}